\documentclass[11pt,reqno]{article}
\usepackage[margin=0.7in]{geometry} 
\usepackage[colorlinks,
            linkcolor=black,
            anchorcolor=black, 
            citecolor=black, 
            urlcolor=blue,
            ]{hyperref}     
\usepackage{indentfirst}
\usepackage{url}            
\usepackage{booktabs}       
\usepackage{amsfonts}       
\usepackage{nicefrac}       
\usepackage{microtype}      
\usepackage[reqno]{amsmath}
\usepackage{bbm}
\usepackage{graphicx}
\usepackage{subcaption}
\usepackage{wrapfig}
\usepackage{enumitem}
\usepackage{multirow}
\usepackage[font=small,labelfont=bf]{caption}
\usepackage{colortbl}
\usepackage[scaled]{helvet}
\usepackage{courier} 
\usepackage{lineno}
\usepackage{bbm}
\usepackage{color}
\usepackage{xcolor}
\usepackage{ulem}
\usepackage{nicematrix}
\usepackage{xspace}

\definecolor{aliceblue}{RGB}{178, 217, 245}

\definecolor{babyblue}{RGB}{217, 239, 251}

\usepackage[fixed]{fontawesome5}

\newcommand{\superscript}[1]{\ensuremath{^\textrm{\footnotesize{#1}}}}
\long\def\symbolfootnote[#1]#2{\begingroup%
\def\thefootnote{\fnsymbol{footnote}}\footnote[#1]{#2}\endgroup}

\usepackage{fontspec}
\title{\vspace{-0.5cm}\Large{\bf{InstructCell: Empowering Single-Cell Analysis by Following multi-modal Instructions}}}

\title{\vspace{-0.5cm}\Large{\bf{OceanPile: A Large-Scale Multimodal Ocean Corpus for Foundation Models}}}
\author{
 \renewcommand{\thefootnote}{\footnotesize{\arabic{footnote}}}
 \bf{Yida Xue}\superscript{1,3}$^{\footnotesize{*}}$,
  \bf{Ningyu Zhang}\superscript{1,3,5}$^{\footnotesize{*}}$$^{\footnotesize{\dag}}$, 
 \bf{Tingwei Wu}\superscript{1,3}$^{\footnotesize{*}}$,
  \bf{Zhe Ma}\superscript{1}$^{\footnotesize{*}}$,\\
 \bf{Daxiong Ji}\superscript{3},
\bf{Zhao Wang}\superscript{3},
 \bf{Guozhou Zheng}\superscript{5,6}$^{\footnotesize{\dag}}$,
 and
 \bf{Huajun Chen}\superscript{1,2,5}
}

\date{}

\begin{document}
\maketitle

\vspace{1cm}

{\renewcommand\baselinestretch{1.4}\selectfont

\noindent $^1$  College of Computer Science and Technology, Zhejiang University, Hangzhou 310027, China. \\
$^2$ ZJU-Hangzhou Global Scientific and Technological Innovation Center, Hangzhou 311200, China. \\
$^3$ School of Software Technology, Zhejiang University, Ningbo 315048, China. \\
$^4$ Ocean College, Zhejiang University, Zhoushan 316021, China. \\
$^5$ State Key Laboratory of Ocean Sensing,
Hangzhou 311200, China. \\
$^6$ Ocean Research Center of  Zhoushan, Zhejiang University, Zhoushan 316021, China. \\

\noindent $^*$ Equal contribution 

\noindent $\dag$ Corresponding Author: Prof. Ningyu Zhang, e-mail: \href{zhangningyu@zju.edu.cn}{\textcolor[RGB]{42, 97, 187}{\uline{zhangningyu@zju.edu.cn}}}, and Prof. Guozhou Zheng, \\e-mail: \href{guozhou@zju.edu.cn}{\textcolor[RGB]{42, 97, 187}{\uline{guozhou@zju.edu.cn}}}

\par}

\renewcommand{\figurename}{Fig.}

\newcommand{\OceanPile}{\textsc{OceanPile}\xspace}
\newcommand{\OceanCorpus}{\textsc{OceanCorpus}\xspace}
\newcommand{\OceanInstruction}{\textsc{OceanInstruction}\xspace}
\newcommand{\OceanBench}{\textsc{OceanBenchmark}\xspace}

\flushbottom

\normalsize

\section*{Abstract}
The vast and underexplored ocean plays a critical role in regulating global climate and supporting marine biodiversity, yet artificial intelligence has so far delivered limited impact in this domain due to a fundamental data bottleneck. Specifically, ocean data are highly fragmented across disparate sources and inherently exhibit multi-modal, high-noise, and weakly labeled characteristics, lacking unified schemas and semantic alignment. Although Multimodal Large Language Models (MLLMs) have achieved remarkable success in general domains, their application to ocean science remains severely constrained by the absence of large-scale, well-aligned multimodal datasets tailored to marine environments. To bridge this gap, we introduce \OceanPile, a large-scale multimodal corpus designed for ocean foundation models. It comprises three key components: \OceanCorpus, a unified collection integrating sonar data, underwater imagery, marine science visuals, and scientific text from diverse authoritative sources; \OceanInstruction, a high-quality instruction dataset synthesized via a novel pipeline guided by a hierarchical Ocean Concept Knowledge Graph; and \OceanBench, a manually curated evaluation benchmark for rigorous assessment. We establish a multi-stage quality control process to ensure scientific validity and alignment across modalities. Experimental validation demonstrates significant performance improvements for models trained on our data. All datasets are publicly released to advance the field of marine artificial intelligence and empower domain-specific MLLMs.


\section*{Background \& Summary}
The world's oceans, covering over 70\% of the Earth's surface, play a fundamental role in regulating global climate, sustaining biodiversity, and supporting economic activities. Despite their critical importance, a large fraction of marine environments remain unexplored, presenting a vast frontier for scientific discovery and technological innovation \cite{falkowski2012ocean,visbeck2018ocean,ocean23,bodnar2025foundation}.
Over the past decades, advances in ocean observation technologies, scientific records, and new discoveries have generated rich multimodal oceanographic data including sonar measurements, complex oceanographic imagery, and domain-specific scientific imagery. These heterogeneous data streams hold the key to unlocking the ocean's enduring mysteries, offering profound insights into marine ecosystems, underwater resources, and global climate processes \cite{lou2023application,zheng2024marineinst,yang2024langya,huang2025fuxi,aubard2025sonar,sonardata,li2024dualsonar}. Meanwhile, building upon the rapid advancement of Large Language Models (LLMs) \cite{llm_survey, llama2023, vicuna2023, qwen}, Multimodal Large Language Models (MLLMs) \cite{mmlm_survey, gpt4o, llama3.2, miniGPT4, gemini, qwenvl_2.5, minicpm, internvl35} have emerged as powerful frameworks for processing and understanding diverse data modalities. These models typically integrate pre-trained LLMs with vision encoders \cite{clip, vit}, aligning them through extensive image-text pairing datasets, thereby enabling comprehensive cross-modal understanding. However, when applied to ocean science, these general-purpose MLLMs have so far delivered limited impact due to a fundamental data bottleneck, which prevents effective knowledge integration and domain-specific reasoning. Consequently, a number of ocean-specific MLLMs \cite{oceangpt, Marinegpt, NAUTILUS} have emerged, yet most of them target only limited subdomains in terms of both model capabilities and training data, rather than addressing the diverse complexity and interdisciplinary nature of marine science.

The ocean data bottleneck stems from multiple intrinsic challenges. First, ocean data have long been in a highly isolated and fragmented state, spanning scientific literature, engineering reports, and observational instruments, while lacking unified schemas and semantic alignment mechanisms. Second, ocean data inherently exhibit multi-modal, high-noise, and weakly labeled characteristics, from sonar signals and remote sensing imagery to biological observations and textual reports, the distributional disparities across modalities are significant and quality is highly uneven, making it difficult to directly support efficient training and reliable reasoning of large models. This fragmentation is compounded by inherent heterogeneity across modalities, where sonar acoustic signatures, visual features in oceanographic imagery, and technical concepts in scientific texts occupy fundamentally distinct semantic spaces. The resulting modality gap and semantic misalignment prevent effective knowledge integration, while the scarcity of large-scale, aligned multimodal datasets specifically tailored for ocean science severely limits MLLMs' ability to develop the domain-specific reasoning capabilities required for marine intelligence tasks.
Traditional sonar datasets \cite{aubard2025sonar, sonardata, li2024dualsonar} and underwater object image datasets \cite{Wildfish, Wildfish2, CoralVQA} are not designed for MLLM training and require extensive preprocessing. Additionally, many datasets are derived from simulated marine environments \cite{MarineGym, holoocean, oceansim, OceanGym}, yet a substantial gap remains between these controlled simulations and the complexity of real-world ocean conditions. As for datasets used to train ocean-specific large models, single-modal approaches like OceanGPT \cite{oceangpt} lack the multimodal inputs needed for comprehensive ocean understanding, while specialized mutimodal approaches such as MarineGPT \cite{Marinegpt} and NautData \cite{NAUTILUS} focus primarily on underwater scene comprehension, overlooking critical aspects of ocean data analysis and interdisciplinary marine science spanning physical, chemical, and biological domains.

\begin{figure*}[t]
\centering 
\includegraphics[width=0.99\textwidth]{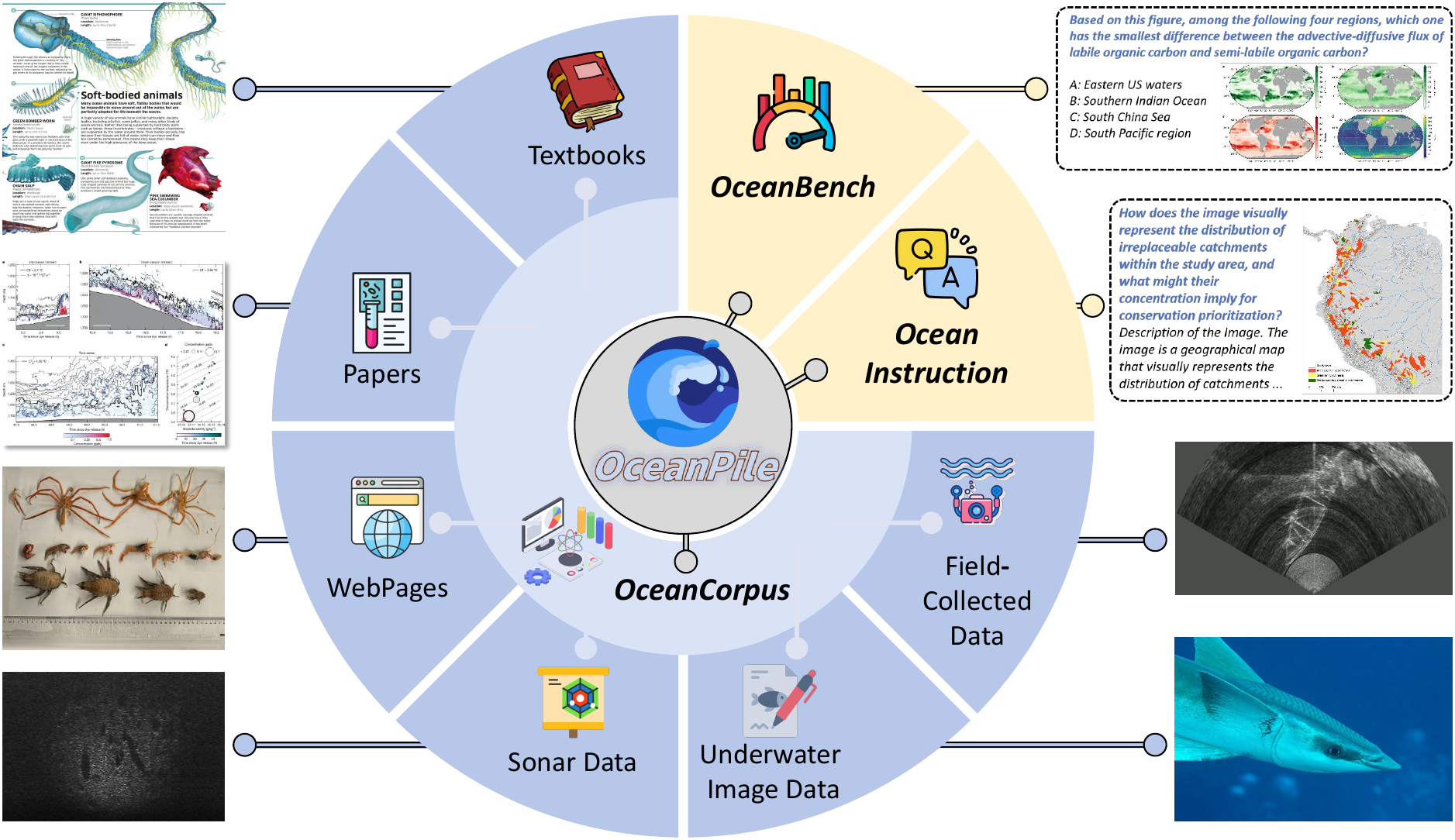}
\caption{
A overview of \OceanPile, which comprises three components: \OceanCorpus, \OceanInstruction, and \OceanBench.
}
\label{fig:intro}
\end{figure*}

To address these critical challenges, we introduce \OceanPile, a large-scale, multimodal corpus specifically designed for ocean foundation models and ocean intelligence. \OceanPile represents the first comprehensive effort to bridge the domain gap in marine AI by providing multi source corpus and carefully aligned multimodal oceanographic data. As illustrated in Fig~\ref{fig:intro}, \OceanPile systematically integrates dispersed and heterogeneous data sources into a unified, open-access resource designed for pre-training with \OceanCorpus, instruction tuning with \OceanInstruction, and evaluation with \OceanBench, thereby creating the essential substrate for developing capable, domain-specific MLLMs \cite{huo2025continuelearningmeetsmultimodal, yang2025surveyspecializedlargelanguage}.
The construction of \OceanPile addresses three fundamental requirements for effective marine AI development: (1) \textbf{Diverse domain-specific sources}: While existing multimodal datasets primarily focus on general web content, \OceanPile aggregates data from specialized marine sources including scientific literature, processed sonar data, biological imagery, and curated content from authoritative oceanographic papers. (2) \textbf{Large-scale multimodal data}: \OceanPile integrates three key types of data: over 5 billion tokens from the foundational multimodal corpus \OceanCorpus for model pre-training, about 140,000 high-quality domain-specific instruction pairs from \OceanInstruction to support supervised fine-tuning and instruction-following capability development, and 1,469 specialized benchmark evaluation samples from \OceanBench to provide a standardized framework for rigorous assessment.
(3) \textbf{Domain-adapted processing pipeline}: We develop specialized data processing techniques that preserve the scientific integrity and contextual richness of oceanographic information to maintain the complex relationships inherent in marine data.

In summary, our contributions are threefold:
\begin{itemize}
    \item We introduce \OceanPile, a large-scale multimodal corpus specifically designed for ocean science, providing aligned data across sonar, imagery, and text modalities to bridge the critical domain gap in marine AI.
    \item We develop specialized data processing pipline that preserve scientific context and ensure semantic alignment across heterogeneous oceanographic data sources, enabling effective training of domain-specific MLLMs.
    \item We establish comprehensive evaluation benchmarks and demonstrate that \OceanPile significantly enhances MLLMs' performance on marine intelligence tasks.
\end{itemize}

\section*{Methods}

\subsection*{Data Collection}
Existing marine datasets lack comprehensive multimodal alignment and interdisciplinary coverage, limiting their utility for MLLMs. To address this limitation, we introduce the \OceanCorpus as the foundational data collection of \OceanPile. The \OceanCorpus integrates multimodal oceanographic data from diverse authoritative sources to ensure comprehensive coverage and scientific validity.

\paragraph{Oceanographic Textbooks.} Comprehensive textbooks provide the foundational knowledge for ocean science. To assemble this corpus, we systematically identify and acquire authoritative oceanography textbooks from major academic publishers and institutional repositories, with the majority of sources in PDF format. Our collection covers diverse subdisciplines including chemical oceanography, biological oceanography, geological oceanography, and physical oceanography.

\paragraph{Oceanographic Papers.} Open-access platforms including ArXiv and Nature portfolio journals provide essential sources of peer-reviewed research spanning diverse marine science disciplines. To construct a representative corpus, we systematically identify relevant publications through a multi-stage filtering approach based on marine-specific keywords, subject categories, and LLM-assisted abstract analysis. Our collection strategy prioritizes acquiring available LaTeX source files while also retaining PDF versions of the papers.

\paragraph{Marine-related Web Pages.}
Online platforms hosting marine-focused content, including scientific news outlets, educational portals, and specialized forums, serve as valuable sources of specialized knowledge about ocean environments and related scientific fields. Our corpus construction systematically gathers content from diverse marine-oriented websites, additionally incorporating high-quality platform links provided by marine experts to ensure the inclusion of authoritative and domain-relevant web resources.

\paragraph{Sonar Detection Datasets.}
We gather specialized publicly available sonar datasets \cite{sonardata, sonardata2, sonardata3} that provide acoustic imaging data obtained through side-scan sonar and multibeam echosounders. Unlike optical imagery, this represents a fundamentally different sensory modality for underwater perception.

\paragraph{Underwater Image Datasets.}
We collect multiple publicly available annotated image datasets \cite{Wildfish, Wildfish2, SCoralDet, CoralVQA} focused on marine biodiversity. These databases contain high-resolution underwater optical images of marine organisms, with each image associated with corresponding labels. Together, they cover diverse species across a wide range of underwater habitats, providing a broad visual foundation for marine species recognition.

\paragraph{Field-Collected Underwater Data.}
To address the limitations of existing datasets, such as restricted object categories and the artificial conditions of laboratory tank-based collections, we deploy autonomous underwater vehicles (AUVs) equipped with both sonar imaging systems and high-resolution optical cameras. These field deployments target the ecologically diverse Chinese Zhoushan marine region, capturing authentic and varied underwater scenes. The collected data includes synchronized sonar images and corresponding optical images, providing a valuable resource for studying marine targets in their natural environments. This new data not only expands the diversity of target classes but also introduces realistic underwater conditions, such as natural lighting variations and complex seabed backgrounds, thereby enhancing the representativeness and robustness of our multimodal corpus.

\subsection*{Data Preprocessing Pipeline}
\paragraph{Preprocessing Pipeline for Textbooks and Papers.}
For source documents, we first process those available in structured formats (e.g., LaTeX or Markdown), directly converting them to clean text while preserving their original logical structure. For documents lacking native structured formats, we employ specialized PDF-to-markdown conversion tools \cite{MinerU} to extract and preserve key content elements including text, images, tables, and hierarchical organization. The conversion process maintains document structure through headings, retains figure and table captions with their contextual descriptions, and ensures accurate encoding of scientific symbols, mathematical formulas, and domain-specific notations.
Following initial conversion, we implement a multi-stage cleaning pipeline. This includes removing peripheral elements such as headers, footers, page numbers, and publication metadata, while applying rule-based filters to eliminate boilerplate text and non-essential reference sections. To enhance content quality and semantic consistency, we utilize large language models for intelligent filtering and semantic deduplication, enabling more nuanced identification of redundant or highly similar content. This approach prioritizes preservation of unique and scientifically substantive marine knowledge while effectively eliminating low-quality and repetitive information. The refined corpus maintains the technical accuracy and contextual richness essential for marine science applications.

\paragraph{Preprocessing Pipeline for Web Pages.}
Web content undergoes a multi-stage cleaning and enhancement pipeline. First, core textual and visual content is extracted using improved HTML parsers that preserve meaningful information while removing non-essential structural elements such as navigation menus, advertisements, and embedded scripts. To refine textual quality, we apply filters informed by established corpus cleaning methodologies, eliminating overly brief, lengthy, or placeholder passages. For associated images, we employ MLLMs to assess visual relevance and quality. Finally, all documents undergo deduplication based on textual similarity.

\paragraph{Preprocessing Pipeline for Target Detection Data.}
To address the heterogeneity and sparse semantics prevalent in sonar and underwater optical image datasets, we implement a standardized preprocessing and annotation enhancement pipeline. Sonar datasets and underwater target detection databases frequently contain annotations in diverse formats and varying levels of detail. We first normalize annotation formats (e.g., converting bounding boxes to a consistent [x1, y1, x2, y2] coordinate system) and align class labels across datasets by merging synonymous categories (e.g., ``cube'' and ``square box''). To enhance the level of detail available in existing sonar datasets, we adopt a two-stage strategy for multimodal annotation enrichment. For data that includes bounding box annotations, we use vision-language models to generate descriptive text that corresponds to specific object locations and their categories. For datasets that only provide image-level labels, the same models are applied to produce textual descriptions focused on broader object categories and overall scene content.
This approach allows us to build an integrated multimodal corpus that ranges from localized, instance-level descriptions to holistic, scene-level representations, thereby improving both the coverage and usability of sonar data.

\begin{figure*}[t]
\centering 
\includegraphics[width=0.99\textwidth]{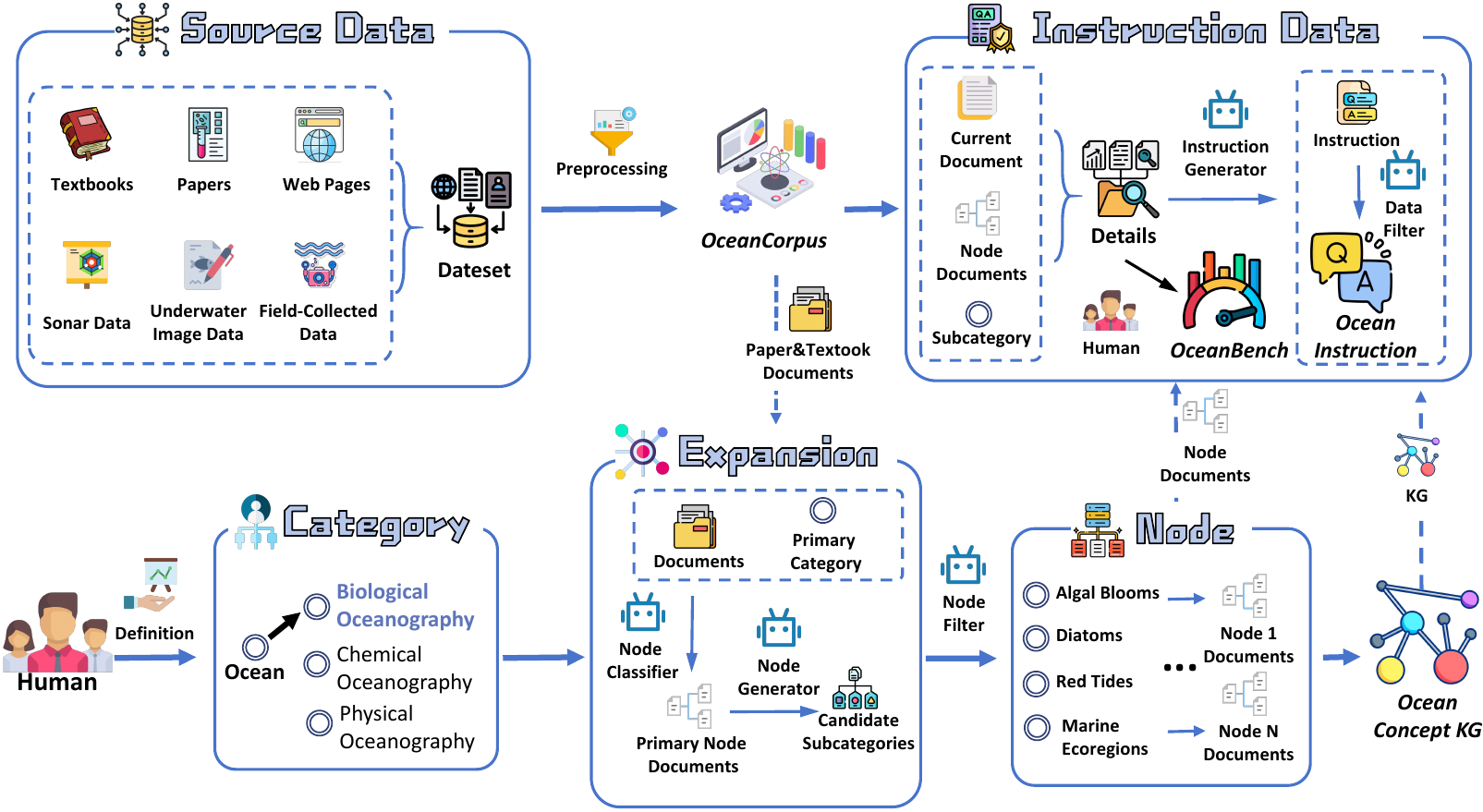}
\caption{
A comprehensive overview of the our framework.
}
\label{fig:framework}
\end{figure*}

\subsection*{Instruction Data Generation}
Instruction data refers to a collection of question-answer pairs designed to train language models to follow specific instructions and perform diverse tasks \cite{zhao2023survey, yin2023survey}. In the context of MLLMs for ocean science, instruction data typically consists of textual prompts paired with desired responses that may incorporate visual or textual information from marine domains. These data enable models to learn how to interpret oceanographic questions, analyze multimodal inputs, and generate scientifically accurate outputs across tasks ranging from simple identification to complex reasoning about marine processes. By providing explicit task descriptions and corresponding ground-truth responses, instruction data bridges the gap between general language understanding and domain-specific competence, allowing MLLMs to develop specialized capabilities for ocean intelligence applications. 

Existing approaches to instruction data generation frequently employ synthetic data methods \cite{survey_synthetic_data} to create instructional content \cite{llava, llavar, LLaVA-CoT, LlamaV-o1}. However, such methods typically exhibit limitations in capturing the depth of scientific knowledge required for specialized domains such as oceanography, and often lack comprehensive coverage of key oceanographic concepts and relationships.
To address these knowledge gaps, we introduce \OceanInstruction, a knowledge-augmented instruction dataset generated through a specialized pipeline for oceanographic domain. As shown in the in Fig~\ref{fig:framework} , our approach begins with the construction of a domain-specific knowledge graph, which is systematically built by extracting and enriching concepts from authoritative scientific literature and structured marine data sources. This foundation enables the generation of instruction-response pairs that are both scientifically accurate and pedagogically structured, ensuring that the resulting data reflects the complexity and interdisciplinary nature of marine science. Under the guidance of the knowledge graph, the instruction data is synthesized and validated.

\paragraph{Ocean Concept Knowledge Graph Construction.}
To systematically structure marine science knowledge and support subsequent instruction synthesis, we construct a hierarchical \textbf{Ocean Concept Knowledge Graph (OCG)}. The construction follows a principled two-phase methodology. Let $\mathcal{D}_{\text{text}}$ denote the corpus of marine textbooks and expert-curated materials, and define the set of primary disciplines (e.g., marine biology, physical oceanography, marine chemistry) as $\mathcal{P} = \{P_k\}_{k=1}^{K}$, where each $P_k$ is identified through consensus from domain experts and textbook taxonomies. For each primary discipline $P_k$, we first extract candidate subcategories $\mathcal{S}_k$ by applying GPT-4o (denoted by $\mathcal{M}$) to the corpus conditioned on $P_k$, yielding
\begin{equation}
\mathcal{S}_k = \mathcal{M}(\mathcal{D}_{\text{text}} \mid P_k).
\end{equation}
We then refine $\mathcal{S}_k$ by employing GPT-4o \cite{gpt4o} to merge similar subcategories and subsequently filter out those with occurrence counts below a threshold $\tau_f$, producing the final set
\begin{equation}
\hat{\mathcal{S}}_k = \mathcal{M}(\mathcal{S}_k; \tau_f).
\end{equation}
This structured representation ensures comprehensive and coherent coverage of core marine science concepts, providing a robust foundation for generating pedagogically sound and scientifically accurate instruction data.

\paragraph{Instruction Data Synthesis.}
Building upon the structured OCG, we generate multimodal instruction data. Let $\mathcal{X}$ denote the input data, which may be a text document $D_j$, a visual element $V_j$ with its description or label $T_j$. For each input $\mathcal{X}_i$, we map it to the most relevant primary discipline $P_k$ and its associated refined subcategory $\hat{\mathcal{S}}_k$ within the OCG, and retrieve pertinent external knowledge $\mathcal{K}_i \subseteq \mathcal{K}$, where $\mathcal{K} = \{K_m\}_{m=1}^l$ is a collection of authoritative sources. The instruction synthesis process is governed by a unified function, $\mathcal{M}$, implemented using GPT-4o:
\begin{equation}
I_i = \mathcal{M}\left(\mathcal{X}_i, P_k, \hat{\mathcal{S}}_k, \mathcal{K}_i\right).
\end{equation}
This function processes the input $\mathcal{X}_i$, enriches it with the structured context of the primary discipline $P_k$, the specific subcategory $\hat{\mathcal{S}}_k$, and supplementary documents $\mathcal{K}_i$, and produces an instruction-answer pair $I_i = (q_i, a_i)$. This unified synthesis pipeline supports three major categories of instruction generation: \textbf{textual data}, \textbf{visual data}, and \textbf{task-specific data}. For \textbf{textual data} such as textbooks and research papers, the model formulates questions that probe key concepts, ensuring foundational knowledge coverage. For \textbf{visual data}, including diagrams and images, it creates queries focused on visual interpretation and scientific description. For \textbf{task-specific data} such as detection-labeled underwater images, it generates instructions tailored to specialized applications like species identification or object analysis, thereby providing comprehensive multimodal training data for marine MLLMs.

\paragraph{Quality Control.}
To ensure the reliability of our generated instruction data, we implement a rigorous multi-stage quality control pipeline. The first stage employs multiple MLLMs as verification agents. Given a source document $\mathcal{X}_i$, a generated question $q_i$, and its corresponding answer $a_i$, each verification agent $\mathcal{V}_j$ (where $j = 1, 2, ..., N$) assigns a quality score $s_{ij} \in [0, 10]$ based on criteria including factual correctness, relevance, and clarity. The final verification score $S_i$ for each instruction-answer pair $I_i = (q_i, a_i)$ is computed as the average across all agents:
\begin{equation}
S_i = \frac{1}{N} \sum_{j=1}^{N} s_{ij},
\end{equation}
and pairs with $S_i$ below a predefined quality threshold $\tau_q$ are automatically filtered out.

Subsequently, we engage domain experts for manual verification. A dedicated platform is developed to allow experts to randomly sample instances from the filtered instruction dataset. Trained marine science experts then meticulously review these sampled instances to identify and correct any remaining errors, ambiguities, or inaccuracies. The consistency of expert judgments is quantified, yielding a high final inter-annotator agreement (IAA) score of 0.86, which indicates strong reliability for research purposes.

\subsection*{OceanBench Construction}
\OceanBench is meticulously constructed as a high-quality evaluation suite for marine MLLMs, comprising two specialized sub-benchmarks: \textbf{Textual Benchmark} for text-only comprehension and \textbf{Multimodal Benchmark} for multimodal reasoning.  The construction follows a rigorous two-stage expert-driven process. First, authoritative marine science documents and aligned multimodal samples are selected. Marine science professionals then design multiple-choice questions based on this curated content.
To ensure benchmark quality, we employ a consensus-driven validation strategy. Each question-answer pair is independently evaluated by $M$ annotators. Let $c_m \in \{0, 1\}$ denote the judgment of the $m$-th annotator, where $c_m = 1$ indicates that the annotator considers the pair to be correct. A pair is retained in the final benchmark only if it receives a majority of positive judgments, formally satisfying the condition

\begin{equation}
\sum_{m=1}^{M} c_m \geq \left\lfloor \frac{M}{2} \right\rfloor + 1.
\end{equation}

This approach ensures the benchmark's high quality through collective expert judgment and majority voting.

\begin{figure*}[t]
\centering 
\includegraphics[width=0.99\textwidth]{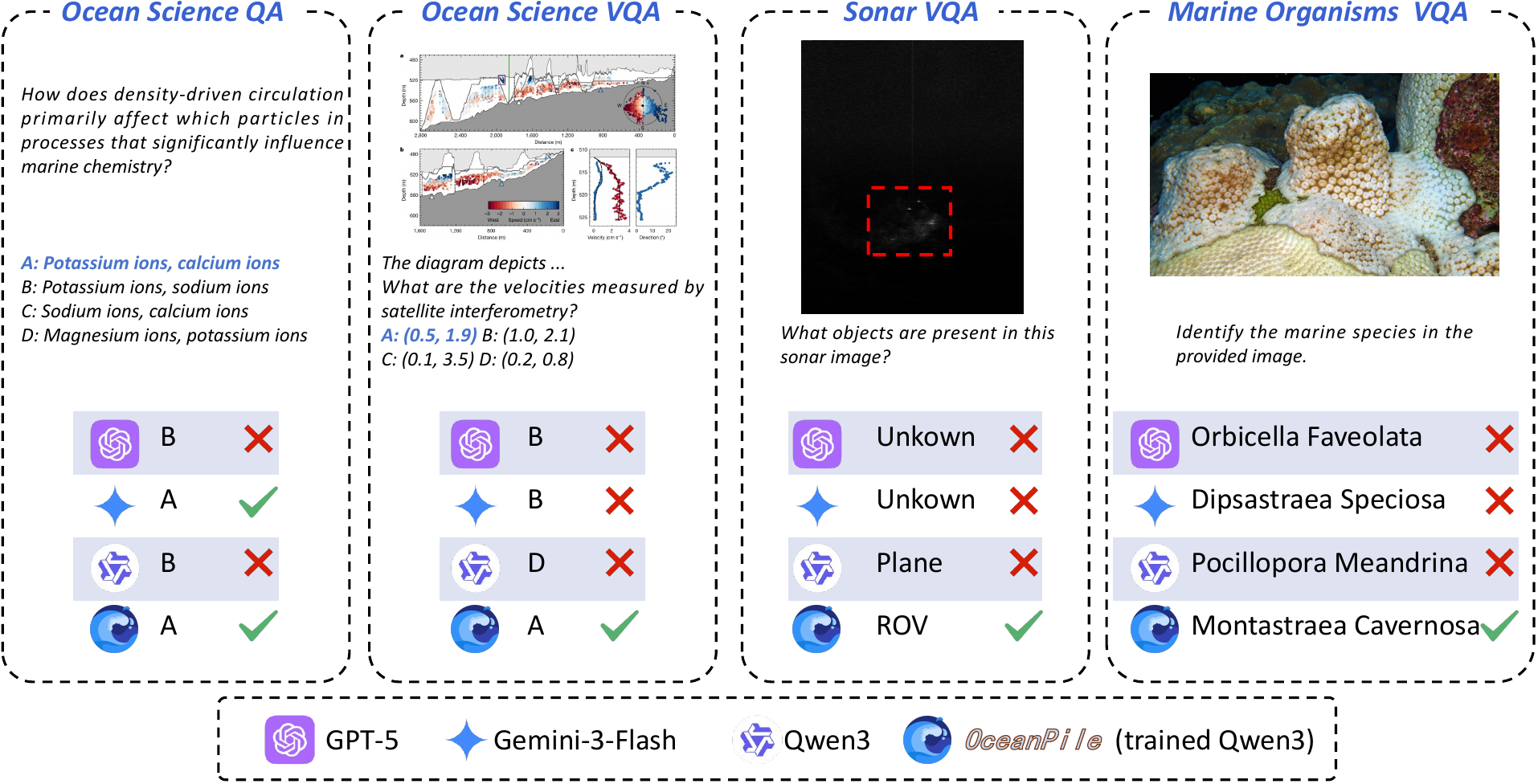}
\caption{
Case analysis.
}
\label{fig:case}
\end{figure*}

\section*{Data Records}
\OceanPile is publicly available on Huggingface at \url{https://huggingface.co/collections/zjunlp/oceanpile}. The repository contains three distinct resources: \OceanCorpus, \OceanInstruction, and \OceanBench. These resources are structured to facilitate research and development of marine-oriented MLLMs.

\paragraph{OceanCorpus.}
\OceanCorpus\ serves as the foundational multimodal data collection. The corpus comprises five main data categories: (1) oceanographic textbooks and papers containing comprehensive marine science knowledge, (2) marine-related web pages providing real-world contextual information, (3) sonar detection databases with acoustic imaging data and annotations, (4) underwater image databases containing optical imagery of marine biodiversity, and (5) field-collected underwater data captured by AUVs in natural marine environments. The corpus preserves both raw source data (over 300,000 PDF documents) and pre-processed multimodal documents (over 5 billion tokens), with the latter stored in CSV format alongside their associated image files, to support flexible data usage.

\paragraph{OceanInstruction.}
\OceanInstruction comprises two distinct versions: text-only instruction dataset, and multimodal instruction dataset. The datasets are provided in structured CSV formats.
Each data instance in both versions includes the following core fields: \texttt{question} (the input query or task description) and \texttt{answer} (the target response). The multimodal instruction dataset additionally includes an \texttt{image} field, which stores the filename or path to the associated visual content (e.g., underwater photographs, scientific diagrams, or sonar images).
The \OceanInstruction (text-only) version contains 69,192 high-quality instruction-answer pairs, while the \OceanInstruction (multimodal) version contains 71,932 instruction-answer pairs, each paired with a relevant marine-themed image. Specifically, the multimodal version includes \textbf{task-specific data} focused on sonar analysis and marine species recognition, as well as \textbf{visual data} centered on marine science. Both versions are designed to provide high-quality, domain-specific instruction data for fine-tuning and evaluating marine-oriented large language models and multimodal models.

\paragraph{OceanBench.}
\OceanBench serves as a comprehensive benchmark for evaluating the marine science capabilities of LLMs and MLLMs. It consists of carefully curated questions spanning both textual and multimodal tasks. The \textbf{Textual Benchmark} focuses on \textbf{Ocean Science QA} (102 samples), assessing models' factual knowledge and reasoning in marine domains through text-only questions. The \textbf{Multimodal Benchmark} is further divided into three specialized sub-benchmarks: \textbf{Ocean Science VQA} (99 samples) evaluates general visual question answering on marine-themed images and diagrams; \textbf{Sonar VQA} (796 samples) targets the interpretation of sonar and acoustic imagery for underwater sensing tasks; and \textbf{Marine Organisms VQA} (472 samples) assesses fine-grained visual recognition and biological knowledge of marine species. Each instance in \OceanBench includes a \texttt{question}, the correct \texttt{answer}, an \texttt{image} (for multimodal tasks), and detailed metadata. The benchmark is provided in CSV format and is intended for testing, enabling standardized evaluation of model performance across both textual and multimodal marine tasks.

\section*{Technical Validation}
\paragraph{Human Verification.}
To ensure the quality and consistency of \OceanInstruction, we implement a multi-round human verification process. After the initial annotation phase, an additional team of independent evaluators conducts a thorough review of the generated instruction-answer pairs. Each instance is assessed by multiple evaluators who independently score its quality based on criteria such as factual accuracy, relevance, clarity, and appropriateness for the marine science domain. Any instance that receives significantly divergent scores (e.g., a standard deviation exceeding a predefined threshold) is flagged for further examination. These flagged instances are then discussed in a consensus meeting involving both the original annotators and the independent evaluators. If a consensus cannot be reached, the instance is removed from the dataset. This rigorous process ensures that only high-quality, unambiguous data is retained, thereby enhancing the overall reliability of the \OceanInstruction dataset.

\paragraph{Model Performance Improvement.}
To quantitatively assess the effectiveness of our instruction data, we fine-tune baseline models including Qwen3-30B-A3B-Instruct and Qwen3-VL-8B-Instruct \cite{qwen3} using \OceanInstruction. Additionally, we evaluate closed-source MLLMs including Gemini-3-Flash \cite{gemini}, GPT-4o\cite{gpt4o}, and  GPT-5 \cite{GPT51} on the \OceanBench. The results, summarized in Table~\ref{tab:mmoceanbench_results}, demonstrate that models trained with our instruction data show significant improvements across both textual and multimodal marine science tasks. To determine correctness, we employ an LLM-as-a-Judge \cite{Gu2024ASO} to compare each model's output against the corresponding ground truth answer. As shown in Figure~\ref{fig:case}, the case examples from each task show that the fine-tuned model can correctly answer ocean science questions and achieves improvements across all tasks.

On the \textbf{Textual Benchmark} (Ocean Science QA), fine-tuning with OceanPile improves Qwen3-30B-A3B-Instruct from 25.49 to 26.47, outperforming GPT-5 (16.67), GPT-4o (6.86), and closely approaching Gemini-3-Flash (24.51). More notably, on the \textbf{Multimodal Benchmark}, Qwen3-VL-8B-Instruct fine-tuned with OceanPile achieves substantial gains across all three sub-benchmarks: from 21.21 to 29.29 on Ocean Science VQA, from 8.04 to 19.97 on Sonar VQA, and from 9.96 to 48.52 on Marine Organisms VQA. This leads to an overall score of 32.59, surpassing GPT-5 (9.67), GPT-4o (14.35), and even outperforming Gemini-3-Flash (31.21) in the overall multimodal evaluation.
These results validate the quality of our instruction data and underscore its utility in advancing the capabilities of LLMs and MLLMs in the marine science domain. The consistent improvements across all evaluated subdomains demonstrate that OceanInstruct effectively bridges the domain adaptation gap for foundation models in marine science.

\begin{table}[htbp]
\centering
\scalebox{0.8}{
\begin{tabular}{llllll}
\toprule
\multirow{2}{*}{\textbf{Model}} & \textbf{Textual Benchmark} & \multicolumn{4}{c}{\textbf{Multimodal Benchmark}} \\
\cmidrule(lr){2-2} \cmidrule(lr){3-6}
& \textbf{Ocean Science QA (\%)} & \textbf{Ocean Science VQA (\%)} & \textbf{Sonar VQA (\%)} & \textbf{Marine Organisms VQA (\%)} & \textbf{Overall (\%)} \\
\midrule
Qwen3-30B & \underline{25.49} & - & - & - & - \\
Qwen3-30B (with OceanPile) & \textbf{26.47}\textcolor{red}{\ensuremath{^{\uparrow 0.98}}} & - & - & - & - \\
Qwen3-VL-8B & - & 21.21 & 8.04 & 9.96 & 13.07 \\
Qwen3-VL-8B (with OceanPile) & - & \underline{29.29}\textcolor{red}{\ensuremath{^{\uparrow 8.08}}} & \textbf{19.97}\textcolor{red}{\ensuremath{^{\uparrow 11.93}}} & \underline{48.52}\textcolor{red}{\ensuremath{^{\uparrow 38.56}}} & \textbf{32.59}\textcolor{red}{\ensuremath{^{\uparrow 19.52}}} \\
GPT-5 & 16.67 & 19.19 & 0.71 & 9.11 & 9.67 \\
GPT-4o & 6.86 & 16.16 & 5.71 & 21.19 & 14.35 \\
Gemini-3-Flash & 24.51 & \textbf{32.32} & \underline{11.11} & \textbf{50.21} & \underline{31.21} \\
\bottomrule
\end{tabular}
}
\caption{Performance comparison of different models on \OceanBench.}
\label{tab:mmoceanbench_results}
\end{table}

\section*{Code Availability}
Related code is available in our Github project: \url{https://github.com/OceanGPT/OceanPile}. Our datasets and models are available on HuggingFace: \url{https://huggingface.co/collections/zjunlp/oceanpile}. For more information, please visit our homepage: \url{http://data.oceangpt.blue/en/}.




\section*{Acknowledgements}
This work was supported by the National Natural Science Foundation of China (Grant No. 62576307) and the Yongjiang Talent Introduction Programme (Grant No. 2021A-156-G). We thank all the researchers and contributors who provided valuable support for this work.

\section*{Author Contributions}
Yida Xue, Tingwei Wu, and Zhe Ma completed the manuscript, data, algorithmic framework, and experiments under the supervision of Ningyu Zhang and Huajun Chen. Ningyu Zhang, Daxiong Ji, Zhao Wang, Guozhou Zheng, and Huajun Chen provided help in methodology design, computing power, storage resources, and marine equipment.

\section*{Competing Interests}
The authors declare no competing interests.




\renewcommand\thefigure{\thesection} 
\renewcommand\thetable{\thesection}    

\bibliographystyle{nature}
\bibliography{ic}

\end{document}